\documentclass[aps, prd, superscriptaddress]{revtex4-2}

\usepackage{amsmath,amssymb,bm}
\usepackage[colorlinks]{hyperref}
\usepackage[capitalise]{cleveref}
\usepackage[dvipdfmx]{graphicx}
\usepackage{multirow}
\usepackage{ulem}
\usepackage{enumitem}
\usepackage{color}

\begin{document}

\title{Realistic assessment of a single gravitational wave source localization taking into account precise pulsar distances with pulsar timing arrays}

\author{Ryo Kato}
\email{ryo.kato@nao.ac.jp}
\affiliation{Mizusawa VLBI Observatory, National Astronomical Observatory of Japan, 2-21-1 Osawa, Mitaka, Tokyo 181-8588, Japan}
\author{Keitaro Takahashi}
\email{keitaro@kumamoto-u.ac.jp}
\affiliation{Faculty of Advanced Science and Technology, Kumamoto University, Kumamoto 860-8555, Japan}
\affiliation{International Research Organization for Advanced Science and Technology, Kumamoto University, Kumamoto 860-8555, Japan
}

\date{\today}

\begin{abstract}
    Pulsar timing arrays (PTAs) are anticipated to detect continuous gravitational waves (GWs) from individual supermassive black hole binaries (SMBHBs) in the near future.
    To identify the host galaxy of a GW source, PTAs require significantly improved angular resolution beyond the typical range of 100-1000 square degrees achieved by recent continuous GW searches.
    In this study, we investigate how precise pulsar distance measurements can enhance the localization of a single GW source.
    Accurate distance information, comparable to or better than the GW wavelength (typically 1~pc) can refine GW source localization.
    In the near future, with the advent of Square Kilometre Array (SKA), such high-precision distance measurements will be feasible for a few nearby pulsars.
    We focus on the relatively nearby pulsars J0437-4715 (156 pc) and J0030+0451 (331 pc), incorporating their actual distance uncertainties based on current VLBI measurements and the anticipated precision of the SKA-era.
    By simulating 87 pulsars with the GW signal and Gaussian white noise in the timing residuals, we assess the impact of the pulsar distance information on GW source localization.
    Our results show that without precise distance information, localization remains insufficient to identify host galaxies under 10 ns noise.
    However, incorporating SKA-era distance precision for nearby pulsars J0437-4715 and J0030+0451 can reduce localization uncertainties to the required level of $10^{-3}$ $\rm deg^{2}$ under 10 ns noise.
    Localization accuracy strongly depends on the geometric configuration of pulsars with well-measured distances and improves notably near and between such pulsars.
    The improvement of the localization will greatly aid in identifying the host galaxy of a GW source and constructing an SMBHB catalog.
    It will further enable follow-up electromagnetic observations to investigate the SMBHB in greater detail.
\end{abstract}

\maketitle

\section{Introduction}\label{sec: Introduction}

A Pulsar Timing Array (PTA) experiment employs high-precision timing of millisecond pulsars to search for gravitational waves (GWs) \cite{fosterConstructingPulsarTiming1990}.
Current PTA experiments include the European Pulsar Timing Array (EPTA) \cite{antoniadisSecondDataRelease2023},
the North American Nanohertz Observatory for Gravitational Waves (NANOGrav) \cite{agazieNANOGrav15Yr2023},
the Parkes Pulsar Timing Array (PPTA) \cite{zicParkesPulsarTiming2023},
the Indian Pulsar Timing Array (InPTA) \cite{tarafdarIndianPulsarTiming2022},
the Chinese Pulsar Timing Array (CPTA) \cite{xuSearchingNanoHertzStochastic2023} and
the MeerKAT Pulsar Timing Array (MPTA) \cite{milesMeerKATPulsarTiming2025}.
The International Pulsar Timing Array (IPTA) aims to improve the sensitivity by integrating data from multiple PTA experiments \cite{pereraInternationalPulsarTiming2019}.
In June 2023, several PTA experiments (EPTA+InPTA \cite{antoniadisSecondDataRelease2023a}, NANOGrav \cite{agazieNANOGrav15Yr2023a}, PPTA \cite{reardonSearchIsotropicGravitationalwave2023} and CPTA \cite{xuSearchingNanoHertzStochastic2023}) have reported non-negligible evidence of the stochastic gravitational wave (GW) background.
Recently, MPTA has also reported the non-negligible evidence \cite{milesMeerKATPulsarTiming2025a}.
The stochastic GW background is potentially produced by a large number of super massive black hole binaries (SMBHBs), inflation, primordial scalar perturbations, cosmological phase transitions, cosmic strings, and domain walls \cite{afzalNANOGrav15Yr2023}.

Similar to the stochastic GW background from the SMBHBs, the detection of continuous gravitational waves (CGWs) from individual SMBHBs will also be expected \cite{rosadoExpectedPropertiesFirst2015,kelleySingleSourcesLowfrequency2018}.
Among the information obtained by detecting the CGWs, information about the sky location of the GW source is important to identify its host galaxy.
Numerous papers have studied the CGWs with mention of the sky location of the GW source \cite{corbinPulsarTimingArray2010,
    sesanaGravitationalWavesPulsar2010,
    sesanaMeasuringParametersMassive2010,
    leeGravitationalWaveAstronomy2011,
    babakResolvingMultipleSupermassive2012,
    ellisPRACTICALMETHODSCONTINUOUS2012,
    ellisOPTIMALSTRATEGIESCONTINUOUS2012,
    petiteauResolvingMultipleSupermassive2013,
    taylorAcceleratedBayesianModelselection2014,
    wangCOHERENTMETHODDETECTION2014,
    zhuAllskySearchContinuous2014,
    wangCOHERENTNETWORKANALYSIS2015,
    zhuDetectionLocalizationSinglesource2015,
    zhuDetectingNanohertzGravitational2015,
    babakEuropeanPulsarTiming2016,
    hazbounNullstreamPointingPulsar2016,
    madisonVersatileDirectionalSearches2016,
    taylorDETECTINGECCENTRICSUPERMASSIVE2016,
    zhuDetectionLocalizationContinuous2016,
    wangPulsarTimingArray2017,
    wangContinuousGravitationalWave2017,
    bakerResultsInternationalPulsar2019,
    goldsteinAssociatingHostGalaxy2019,
    obeirneConstrainingAlternativePolarization2019,
    becsyJointSearchIsolated2020,
    taylorBrightBinariesBumpy2020,
    songshengSearchContinuousGravitationalwave2021,
    wangExtendingFrequencyReach2021,
    becsyFastBayesianAnalysis2022,
    chenParameterestimationBiasesEccentric2022,
    falxaSearchingContinuousGravitational2023,
    katoPrecisionLocalizationSingle2023,
    antoniadisSecondDataRelease2024,
    becsyEfficientBayesianInference2024,
    ferrantiSeparatingDeterministicStochastic2024,
    kimpsonKalmanTrackingParameter2024,
    kimpsonStatespaceAnalysisContinuous2024,
    petrovIdentifyingHostGalaxies2024,
    furusawaResolvingIndividualSignal2025,
    gardinerCharacterizingContinuousGravitational2025}.
For example, in the CGW analysis of the IPTA Second Data Release \cite{falxaSearchingContinuousGravitational2023}, there is a case where the sky location of the GW source is relatively well constrained, although the uncertainty remains on the order of 100 to 1000 square degrees.
It has also been suggested that the angular separation between the pulsars and the sky location of the CGW signal model is an important factor in interpreting these localization results.

In our previous study \cite{katoPrecisionLocalizationSingle2023}, the effect of precise pulsar distance measurements on the determination of sky location of a single GW source is investigated, and it was found that a few pulsars with accurately measured distances can improve the sky localization by several orders of magnitude.
In that analysis, we assumed that the pulsars were uniformly distributed on the celestial sphere.
However, in reality, pulsars are predominantly concentrated along the Galactic plane.
Based on the IPTA CGW analysis results, the localization accuracy of the CGW source is likely to depend on the relative spatial configuration between the pulsars and the source.
Recent simulation studies have pointed out that, to improve the localization accuracy, a small angular separation between the CGW source and nearby pulsars plays a significant role \cite{petrovIdentifyingHostGalaxies2024}.
Therefore, in this study, we consider the actual spatial distribution of the pulsars used in current pulsar timing arrays.

Furthermore, in our previous study, we emphasized the importance of obtaining pulsar distance information independently from PTAs through very long baseline interferometry (VLBI) measurements, but we uniformly varied the distance uncertainties derived from VLBI measurements for all pulsars.
In reality, pulsar distances vary significantly, and consequently the expected distance uncertainties differ from one pulsar to another.
In this study, we focus on the PSRs J0437-4715 and J0030+0451, which have small distance uncertainties because they are close to Earth.
The distance to J0437-4715 is 156 pc with a current uncertainty of 1.3 pc \cite{dellerPRECISIONSOUTHERNHEMISPHERE2009}, while the distance to J0030+0451 is 331 pc with a current uncertainty of 7.9 pc \cite{hdingMSPSRpCatalogueVLBA2023}.
Future observations with the Square Kilometre Array (SKA) have the potential to achieve an angular precision of 15 $\rm \mu as$ \cite{smitsProspectsAccurateDistance2011}, in which case the distance uncertainties for the two pulsars are expected to be 0.37 pc and 1.7 pc, respectively.

Therefore, compared to our previous study, this study considers more realistic cases for both the spatial distribution of pulsars and their distance uncertainties, which allows for a more realistic evaluation of the localization accuracy of gravitational-wave sources.
Additionally, in our previous analysis, we only considered cases where the timescale of binary evolution was significantly longer than the time difference between the Earth term and the pulsar term.
In this study, we also examine cases where the evolution timescale is comparable to this time difference and assess whether this factor affects localization accuracy.

In this paper, we review a method for computing constant probability density contours of parameters using closed-form (analytical) posterior distributions, which can be used when the data are simulated.
We consider a total of 87 pulsars, resulting in a much larger dataset and a greater number of parameters compared to the 12 pulsars used in our previous study.
In such a situation, running Markov Chain Monte Carlo (MCMC) becomes challenging, so it was necessary to derive the posterior distribution analytically and evaluate it only once.

The paper is organized as follows.
In \cref{sec: Continuous GW Signal}, we review the CGW signal induced by a GW emitted from a circular SMBHB.
In \cref{sec: Simulated Data}, we describe our simulated data.
In \cref{sec: Bayesian influence}, we first introduce Bayesian parameter estimation and discuss the posterior and prior distributions. We then review a method for computing constant probability density contours.
In \cref{sec: Results}, we present our results and assess the impact of precise pulsar distance information on GW source localization.
In \cref{sec: Conclusion}, we present our conclusions and discussions.
Throughout this paper, we use units where $G = c = 1$.

\section{Continuous GW Signal}\label{sec: Continuous GW Signal}

Timing residuals are defined as the differences between the predicted times of arrival (TOAs) of the pulses emitted by the pulsar and the actual TOAs.
Since the GW affects the light propagation time, it induces the timing residuals \cite{finnResponseInterferometricGravitational2009}.
In this paper, we consider a continuous GW emitted by a single SMBHB, called a CGW signal.
In this section, we briefly review the timing residuals induced by a circular SMBHB.

Following the article \cite{sesanaMeasuringParametersMassive2010}, we write the GW as
\begin{align}
    h_{\mu\nu}(t,\hat{\Omega})=\sum_{A=+,\times}h_{A}(t,\hat{\Omega})e_{\mu\nu}^{A}(\hat{\Omega}),
\end{align}
where $\hat{\Omega}$ is a unit vector from the GW source to the Solar System Barycenter (SSB), $h_{A}$ $(A = +, \times)$ are the polarization amplitudes and $e_{\mu\nu}^{A}$ are the polarization tensors.
The polarization tensors are expressed as
\begin{align}
    e_{\mu \nu}^{+}(\hat{\Omega})=\hat{m}_{\mu}\hat{m}_{\nu}-\hat{n}_{\mu}\hat{n}_{\nu}, \\
    e_{\mu \nu}^{\times}(\vec{\Omega})=\hat{m}_{\mu}\hat{n}_{\nu}+\hat{n}_{\mu}\hat{m}_{\nu},
\end{align}
where $\hat{m}$ and $\hat{n}$ are unit vectors orthogonal to each other and to $\hat{\Omega}$.
The unit vectors can be defined as:
\begin{align}
    \hat{\Omega} & =-(\sin\theta\cos\phi)\hat{x}-(\sin\theta\sin\phi)\hat{y}-(\cos\theta)\hat{z}, \\
    \hat{m}      & =(\sin\phi)\hat{x}-(\cos\phi)\hat{y},                                          \\
    \hat{n}      & =-(\cos\theta\cos\phi)\hat{x}-(\cos\theta\sin\phi)\hat{y}+(\sin\theta)\hat{z},
\end{align}
where ${\hat{x}},{\hat{y}}$ and ${\hat{z}}$ are the basis vectors in the SSB Cartesian coordinate system and $\hat{\Omega}=\hat{m}\times\hat{n}$.
The sky location of the GW source is specified by the polar angle $\theta$ and the azimuthal angle $\phi$.
The antenna pattern function of the PTA, which encodes the angular sensitivity of the GW detector, is given by \cite{bookAstrometricEffectsStochastic2011}:
\begin{align}
    F^{A}_{p}({\hat{\Omega}})={\frac{1}{2}}{\frac{{\hat{p}_{p}}^{\mu}{\hat{p}_{p}}^{\nu}}{1+{\hat{\Omega}}\cdot{\hat{p}_{p}}}}e_{\mu\nu}^{A}({\hat{\Omega}}),
\end{align}
where $\hat{p}_{p}$ is a unit vector from the SSB to the pulsar whose sky location is speciﬁed by the polar angle $\theta_{p}$ and the azimuthal angle $\phi_{p}$:
\begin{align}
    \hat{p}_{p}  =(\sin\theta_{p}\cos\phi_{p})\hat{x}+(\sin\theta_{p}\sin\phi_{p})\hat{y}+(\cos\theta_{p})\hat{z}.
\end{align}
Here we will use the index $p$ to label a particular pulsar ($p = 1, \dots, N_{\rm psr}$ where $N_{\rm psr}$ is the number of pulsars).

The CGW induces the timing residuals in the following form \cite{ellisSearchingGravitationalWaves2014}:
\begin{align}
    s_{p}(t,\hat{\Omega})=\sum_{A=+,\times}\Delta s^{A}_{p}(t)F^{A}_{p}(\hat{\Omega}),\label{eq:signal}
\end{align}
where
\begin{align}
    \Delta s^{A}_{p}(t)=s^{A}(t_{p})-s^{A}(t),
    \label{eq:residual}
\end{align}
and $t$ and $t_{p}$ are times when the CGW passed through the SSB and the pulsar, respectively, and $t$ also denotes the time measured on the SSB.
The term associated with $t_{p}$ in the GW-induced timing residuals is called the pulsar term, and the term associated with $t$ is called the Earth term.
The relation between $t$ and $t_{p}$ is as follows
\begin{align}
    t_{p}    & =t-\tau_{p},                                                      \\
    \tau_{p} & \equiv L_{p}(1+{\hat{\Omega}}\cdot{\hat{p}}),\label{timerelation}
\end{align}
where $L_{p}$ is the distance between the SSB and the pulsar.
Specifically, in the case of a CGW emitted from a circular SMBHB, $s^{A}$ can be written as
\begin{align}
    s^{+}(t)      & =\frac{{\mathcal M}^{5/3}}{d_{L}\omega_{s}(t)^{1/3}}\left[\sin2\Phi_{s}(t)\left(1+\cos^{2}\iota\right)\cos2\psi+2\cos2\Phi_{s}(t)\cos\iota\sin2\psi\right],          \label{plussignal}\\
    s^{\times}(t) & =\frac{\mathcal{M}^{\mathrm{5/3}}}{d_{L}\omega_{s}(t)^{1/3}}\left[-\sin2\Phi_{s}(t)\left(1+\cos^{2}\iota\right)\sin2\psi+2\cos2\Phi_{s}(t)\cos\iota\cos2\psi\right],\label{crosssignal}
\end{align}
where $\mathcal{M}\equiv(m_{1}m_{2})^{3/5}/(m_{1}+m_{2})^{1/5}$ is the chirp mass of the SMBHB with the individual black hole masses $m_{1}$ and $m_{2}$, $d_{L}$ is the luminosity distance of the SMBHB, $\iota$ is the inclination angle of the SMBHB and $\psi$ is the GW polarization angle.
The orbital angular frequency and the phase of the SMBHB are
\begin{align}
    \omega_{s}(t) &
    =\omega_{s0}\left(1-\frac{256}{5}{\mathcal M}^{5/3}\omega_{s0}^{8/3}t\right)^{-3/8}
    =\omega_{s0}\left(1-\frac{t}{t_{\rm coal}}\right)^{-3/8}, \label{eq:omega} \\
    \Phi_{s}(t)   &
    =\Phi_{s0}+{\frac{1}{32{\mathcal M}^{5/3}}}\left(\omega_{s0}^{-5/3}-\omega_{s}(t)^{-5/3}\right)
    =\Phi_{s0}+\frac{8}{5} \omega_{s0} t_{\rm coal} \left[1-\left(1-\frac{t}{t_{\rm coal}}\right)^{5/8}\right],\label{eq:phase}
\end{align}
where $\omega_{s0}\equiv2\pi f_{s0}$ and $\Phi_{s0}$ are the initial values of the orbital angular frequency and phase at $t=0$, respectively.
Here, the coalescence time of the SMBHB, at which the orbital angular frequency $\omega_{s}$ diverges, is defined as
\begin{align}
    t_{\rm coal}=\frac{5}{256}\mathcal{M}^{-5/3}\omega_{s0}^{-8/3}.
\end{align}
The initial GW frequency and phase are related to these quantities as $f_{0}=2f_{s0}$ and $\Phi_{0}=2\Phi_{s0}$, since the GW frequency and phase are twice the orbital frequency and phase of SMBHB, respectively.
The coalescence time $t_{\rm coal}$ is often compared to the time difference between the Earth term and the pulsar term $\tau_{p}$ of \cref{timerelation} and the time span of the observations.
If the coalescence time is sufficiently longer than the time difference between the Earth term and the pulsar term, the orbital angular frequency of \cref{eq:omega} and phase of \cref{eq:phase} will be the same in the Earth term and the pulsar term.
Also, if the coalescence time is sufficiently longer than the observation period, the orbital angular frequency and phase will not change within the observation period.

Finally, we define the initial amplitude of the Earth term at $t=0$ as
\begin{align}
    A_{e}=\frac{{\mathcal M}^{5/3}}{d_{L}\omega_{s0}^{1/3}}.
\end{align}

\section{Simulated Data}\label{sec: Simulated Data}
\begin{figure}[tb]
    \centering
    \includegraphics[width=14cm]{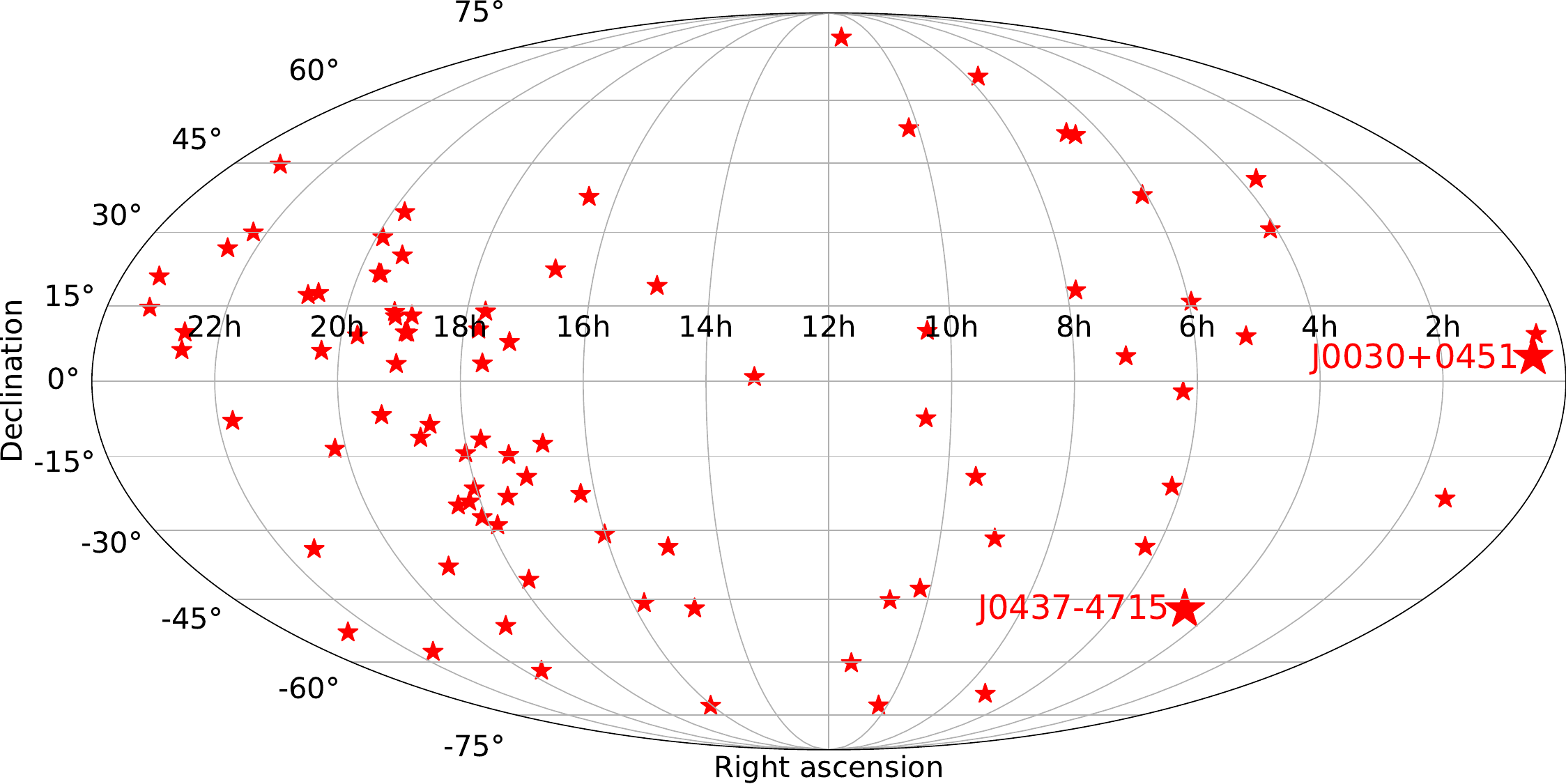}
    \caption{Sky locations of the pulsars, where the star markers denote the sky locations of the pulsars.
        The two pulsars of interest (J0030+0451 and J0437-4715) are highlighted with larger star markers.
    }
    \label{fig:pos}
\end{figure}

For this work, we create a simulated PTA dataset consists of the pulsars with actual sky locations.
The sky locations of the pulsars are extracted from the three datasets of EPTA, NANOGrav and PPTA, which contain a total of 87 pulsars \cite{antoniadisSecondDataRelease2023,
    agazieNANOGrav15Yr2023,
    zicParkesPulsarTiming2023}.
\cref{fig:pos} shows the sky location of the pulsars.
The declination $\delta$ and right ascension $\alpha$ are related to the polar angle $\theta_p$ and azimuthal angle $\phi_p$ by $\delta = \pi/2 - \theta_p$ and $\alpha = \phi_p$, respectively.
All pulsars are assumed to have a distance of 1 kpc.
The time span of the observations is 12.5 years and the observing cadence is three weeks.
The initial value of the observation time is set to zero.

We added the CGW signal and the Gaussian white noise to the dataset.
In the SKA era, a timing precision is less than 100 ns \cite{janssenGravitationalWaveAstronomy2014}.
So we will consider three different standard deviations of the Gaussian white noise: 10, 30, and 100 ns.

To clarify the significance of these noise levels, in current PTAs, the rms white-noise levels for most pulsars are around 1 $\mu$s, with the high-precision pulsar J1909-3744 reaching 140 ns \cite{antoniadisSecondDataRelease2023}.
Consequently, a noise level of 100 ns has not yet been achieved in current PTAs.
Looking ahead to the SKA era, simulations considering only pulse phase jitter noise and radiometer noise indicate that the number of pulsars with rms noise levels of 100 ns is expected to be on the order of 1000, while those with rms noise levels of 30 ns are estimated to number approximately 100 in SKA2-Mid \cite{weltmanFundamentalPhysicsSquare2020}.
Thus, the scenario considered in this study, in which approximately 100 pulsars reach 10 ns, represents an idealized situation, whereas 30 ns is realistic in the SKA era, and 100 ns is expected to be practically achievable.

We will discuss three different CGW signals which add to the dateset listed in \cref{parameters}.
The three CGW signals are distinguished by the coalescence time $t_{\rm coal}$ and are referred to as no, slow, and fast evolution.
In the case of fast evolution, the coalescence time $t_{\rm coal}$ (2050 yr) is not much longer than the delay between the Earth term and the pulsar term $\tau_{p}$ of \cref{timerelation} (up to $2 \times 1$ kpc $\simeq6500$ ly).
Also, in the all cases, the coalescence time is much longer than the observation period (12.5 yr).
The initial amplitude of the Earth term $A_{e}$ is set to the same value for all GW signals.

\begin{table}[htbp]
    \caption{CGW signal parameters for simulated datasets.}
    \label{parameters}
    \centering
    \normalsize
    \begin{center}
        \begin{tabular}{l|lllllllll|ll}
            \hline
            \hline
                           & $\theta$\footnote{We consider various values for GW sky locations ($\theta$ and $\phi$).} & $\phi$\footnotemark[1] & $\iota$ & $\mathcal M$           & $f_{0}$              & $d_{L}$                 & $\Phi_{0}$ & $\psi$ & $L_{p}$ & $A_{e}$         & $t_{\rm coal}$ \\
            \hline
            No evolution   & -                                                                                         & -                      & 1       & $10^{8.2}$ $M_{\odot}$ & $10^{-8}$ ${\rm Hz}$ & $10^{0.67}$ ${\rm Mpc}$ & 1          & 1      & 1 kpc   & $44$ ${\rm ns}$ & 951650 yr      \\
            Slow evolution & -                                                                                         & -                      & 1       & $10^9$ $M_{\odot}$     & $10^{-8}$ ${\rm Hz}$ & $10^2$ ${\rm Mpc}$      & 1          & 1      & 1 kpc   & $44$ ${\rm ns}$ & 44172 yr       \\
            Fast evolution & -                                                                                         & -                      & 1       & $10^{9.8}$ $M_{\odot}$ & $10^{-8}$ ${\rm Hz}$ & $10^{3.33}$ ${\rm Mpc}$ & 1          & 1      & 1 kpc   & $44$ ${\rm ns}$ & 2050 yr        \\
            \hline
            \hline
        \end{tabular}
    \end{center}
\end{table}

Finally, we discuss the amplitudes of CGW signals that are currently detectable.
For the slow-evolution case, we adopt $\mathcal{M} = 10^9 \, M_{\odot}$, $f_0 = 10^{-8} \, {\rm Hz}$, and $d_L = 10^2 \, {\rm Mpc}$.
For the same mass and frequency, current PTA limits place a lower bound on the distance of approximately 10 Mpc \cite{antoniadisSecondDataRelease2024}.
Consequently, the slow-evolution case corresponds to a signal whose amplitude is about one-tenth of the current observational limit, since the amplitude of a CGW signal scales inversely with distance, as given by \mbox{\cref{plussignal,crosssignal}}.
Thus, the amplitude is below the detection threshold of current PTAs.
Nevertheless, it is sufficiently large to be considered realistic for simulations of future high-precision PTA datasets.

\section{Bayesian influence}\label{sec: Bayesian influence}
\subsection{Parameter Estimation}\label{subsec: Parameter Estimation}

In the Bayesian framework, the posterior probability density function for parameters $\vec \theta$ is given by \cite{gregoryBayesianLogicalData2005}:
\begin{align}
    p(\vec \theta|\bm{\delta t},M)=\frac{p(\bm{\delta t}|\vec \theta,M)p(\vec \theta|M)}{p(\bm{\delta t}|M)},\label{posterior}
\end{align}
where $M$ is a parameterized model, $\bm{\delta t}$ is observed timing residuals has a dimension equal to the number of TOAs, $p(\bm{\delta t}|\vec \theta,M)$ is the likelihood function, $p(\vec \theta|M)$ is the prior probability density function and $p(\bm{\delta t}|M)$ is a normalization constant defined as:
\begin{align}
    p(\bm{\delta t}|M)=\int d\vec\theta\,p(\bm{\delta t}|\vec \theta,M)p(\vec \theta|M).\label{evidence}
\end{align}
The goal of parameter estimation is to obtain the posterior distribution.

Given a model $M$ contains the CGW signal model $\bm{s}(\vec \theta)$ and the Gaussian white noise $\bm{n}$, the likelihood function can be written as \cite{gregoryBayesianLogicalData2005}:
\begin{align}
    p(\bm{\delta t}|\vec \theta,M)=\frac{1}{\sqrt{2\pi \bm{C}}}\exp\left(-\frac{1}{2}(\bm{\delta t}-\bm{s}(\vec \theta))^{T}\bm{C}^{-1}(\bm{\delta t}-\bm{s}(\vec \theta))\right),\label{likelihood}
\end{align}
where $\vec \theta$ denotes the parameters for the CGW signal model, $\bm{C} \equiv \langle \bm{n} \bm{n}^T \rangle_{n}$ is the covariance matrix of the Gaussian white noise with a zero mean $\langle \bm{n}\rangle_{n}=0$, and $\langle ... \rangle_{n}$ denotes integration over the probability distribution of the noise.

The covariance matrix $\bm{C}$ is proportional to the identity matrix $\bm{I}$ as
\begin{align}
    \bm{C}=\sigma_{n}^{2} \bm{I},
\end{align}
where $\sigma_{n}$ is a standard deviation of the Gaussian white noise.

We define the specific expression of the timing residuals as the concatenation of the timing residuals of each pulsar:
\begin{align}
    \bm{\delta t}=(\bm{\delta t}_{1},\bm{\delta t}_{2},\cdots,\bm{\delta t}_{N_{\rm psr}})^{T},
\end{align}
and the timing residuals of each pulsar is
\begin{align}
    \bm{\delta t}_{p} & =(\delta t_{p}^{1},\delta t_{p}^{2},\cdots,\delta t^{N_{\rm TOA}}_{p})^{T},
\end{align}
where $N_{\rm TOA}$ is the number of the TOAs of a pulsar, and each timing residual is observed at each TOA.
According to this definition, the CGW signal model is also defined as follows:
\begin{align}
    \bm{s}(\vec \theta) & =(\bm{s}_{1},\bm{s}_{2},\cdots,\bm{s}_{N_{\rm psr}})^{T},                                              \\
    \bm{s_{p}}          & =(s_{p}(t_{1},\hat{\Omega}),s_{p}(t_{2},\hat{\Omega}),\cdots,s_{p}(t_{N_{\rm TOA}},\hat{\Omega}))^{T},
\end{align}
where $s_{p}(t_{a},\hat{\Omega})$ of \cref{eq:signal} is observed at a specific TOA, and $a = 1, \dots, N_{\rm TOA}$.
We also define that all of the noise for all of the pulsars and TOAs are characterized by the same Gaussian white noise.

\subsection{Closed Form Posterior}\label{subsec: Closed Form Posterior}

In this section, we derive the closed-form expression of the posterior distribution for both uniform and Gaussian priors.
Specifically, we show that the integral of the normalization constant in \cref{evidence} of the posterior distribution can be evaluated and expressed in a closed-form (i.e., analytical) expression.
We refer to this as the closed-form posterior distribution.
This distribution can be computed when the true signal and noise are known, that is, when the data are artificially generated.
It can be used to assess the performance of parameter estimation.
Once a closed-form expression is obtained, the posterior distribution can be computed without relying on numerical Bayesian sampling methods such as MCMC.

We consider that the observed timing residuals $\bm{\delta t}$ consist of the true CGW signal $\bm{s}(\vec\theta_0)$ and the Gaussian white noise $\bm{n}$:
\begin{align}
    \bm{\delta t}=\bm{s}(\vec\theta_0)+\bm{n}, \label{true data}
\end{align}
where $\vec\theta_0$ denotes the true value of the CGW parameters.
Then, we use a model $M$ in which the data are represented as the sum of the CGW signal model $\bm{s}(\vec\theta)$ and Gaussian white noise $\bm{n}$.

In the high signal-to-noise ratio (SNR) regime, the posterior distribution is expected to be sharply peaked around the true value of the parameter.
Follwing articles \cite{vallisneriUseAbuseFisher2008,rodriguezInadequaciesFisherInformation2013}, we expand the CGW signal model $\bm{s}(\vec \theta)$ to first order around the true parameter value $\vec\theta_0$ as
\begin{align}
    \bm{s}(\vec\theta) & \simeq\bm{s}(\vec\theta_0)+\Delta\theta_{i}\bm{s}_{i}, \label{linear model}                              \\
    \Delta\theta_{i}   & \equiv \theta_{i}-\theta_{0i},                                                                      \\
    \bm{s}_{i}         & \equiv\left.\frac{\partial}{\partial\theta_{i}}\bm{s}(\vec\theta)\right|_{\vec\theta=\vec\theta_0},\label{diff gw}
\end{align}
where index $i$ denotes the component of the vector for the parameter and the Einstein summation convention is used for Latin indices throughout this section.
The previous article has shown that the posterior distribution obtained using a linearized waveform corresponds to the leading-order term in an expansion of the posterior distribution in powers of the inverse of the optimal SNR of the true signal \cite{vallisneriUseAbuseFisher2008}.
In this context, the optimal SNR of the true signal represents the total SNR of the dataset, as defined in \cref{snr} in Appendix, but it is evaluated using the true parameter values of the CGW signal model rather than the maximum-likelihood estimates.
Follwing articles \cite{vallisneriUseAbuseFisher2008,rodriguezInadequaciesFisherInformation2013}, we refer to this limit as the high signal-to-noise ratio / linearized-signal approximation (high-SNR/LSA) limit.
To compute the above function, the true values of the CGW signal model must be known.
However, even if the true values are known, the shift of the peak of the posterior distribution from the true values due to noise remains unknown.

Substituting \cref{linear model,true data} into the likelihood function of \cref{likelihood} and expanding the exponent in powers of $\vec\theta$, we obtain
\begin{align}
    p(\bm{\delta t}|\vec \theta,M) & =\frac{1}{\sqrt{\det(2\pi \bm{C})}}\exp\left[-\frac{1}{2}\left(\theta_{i}H_{ij}\theta_{j}-2H_{ij}\left(\theta_{0j}+\left(H^{-1}\right)_{jk}n_{k}\right)\theta_{i}+c\right)\right], \\
    n_{i}                          & \equiv \bm{n}^{T}\bm{C}^{-1}\bm{s}_{i},                                                                                                                                            \\
    H_{ij}                         & \equiv \bm{s}^{T}_{i}\bm{C}^{-1}\bm{s}_{j},                                                                                                                                        \\
    c                              & \equiv2n_{i}\theta_{0i}+\theta_{0i}H_{ij}\theta_{0j}+\bm{n}^{T}\bm{C}^{-1}\bm{n},
\end{align}
where we assume that a symmetric matrix $H$, whose comoponent is $H_{ij}$, is a positive definite matrix.
In order to obtain a closed form of the posterior distribution, we need to integrate the normalization constant of \cref{evidence}.
We will consider uniform and Gaussian distributions for the prior so that we can use the following Gaussian integral formula \cite{zinn-justinPathIntegralsQuantum2010}:
\begin{align}
    \int d\vec\theta\exp\left[-\frac{1}{2}\left(\theta_{i}A_{ij}\theta_{j}-2b_{i}\theta_{i}\right)\right]=\sqrt{\det(2\pi A^{-1})}\exp\left[\frac{1}{2}b_{i}\left(A^{-1}\right)_{ij}b_{j}\right],\label{gauss}
\end{align}
where $A$ is a symmetric positive definite matrix.
It is clear that the covariance of the noise $\bm{C}$ cannot be parametrized if we want to use the Gaussian integral formula.

First, we consider a uniform prior $p(\vec \theta|M)=1$.
Using the Gaussian integral formula of \cref{gauss}, the normalization constant of \cref{evidence} is
\begin{align}
    p(\bm{\delta t}|M) & =\int d\vec\theta\,p(\bm{\delta t}|\vec \theta,M),\notag                                                                                                                                              \\
                       & =\frac{\sqrt{\det(2\pi H^{-1})}}{\sqrt{\det(2\pi \bm{C})}}\exp\left[\frac{1}{2}\left(\left(\theta_{0i}+\left(H^{-1}\right)_{ik}n_{k}\right)H_{ij}\left(\theta_{0j}+\left(H^{-1}\right)_{jl}n_{l}\right)-c\right)\right].
\end{align}
Then the posterior distribution of \cref{posterior} can be written as:
\begin{align}
    p(\vec \theta|\bm{\delta t},M) & =\frac{1}{\sqrt{\det(2\pi\Sigma)}}\exp\left[-\frac{1}{2}\left(\left(\theta_{i}-\mu_{i}\right)\left(\Sigma^{-1}\right)_{ij}\left(\theta_{j}-\mu_{j}\right)\right)\right], \\
    \mu_{i}                        & \equiv\langle \theta_{i}\rangle=\theta_{0i}+\left(H^{-1}\right)_{ij}n_{j},                                                                                               \\
    \Sigma_{ij}                    & \equiv\langle(\theta_{i}-\langle\theta_{i}\rangle)(\theta_{j}-\langle\theta_{j}\rangle)\rangle=\left(H^{-1}\right)_{ij},\label{uniform}
\end{align}
where we define the expected value of the function $f(\vec\theta)$ of the parameter as follows
\begin{align}
    \langle f(\vec\theta)\rangle\equiv\int d\vec\theta\,f(\vec\theta) p(\vec \theta|\bm{\delta t},M). \label{bayes expected value}
\end{align}
It can be seen that the posterior distribution is a multivariate normal distribution.

We also consider a Gaussian prior with a mean $\theta_{0i}$ and a covariance $D$ as:
\begin{align}
    p(\vec\theta|M) & =\frac{1}{\sqrt{\det(2\pi D)}}\exp\left[-\frac{1}{2}\left(\theta_{i}-\theta_{0i}\right)\left(D^{-1}\right)_{ij}\left(\theta_{j}-\theta_{0j}\right)\right] \quad\text{(Gaussian prior)}.\label{gaussian prior}
\end{align}
Using the same procedure as in the uniform prior case, it is found that the covariance of the posterior only changes from $H$ to $H+D^{-1}$ compared to the uniform prior case.
Specifically, the posterior distribution with the Gaussian prior is
\begin{align}
    p(\vec \theta|\bm{\delta t},M) & =\frac{1}{\sqrt{\det(2\pi\Sigma)}}\exp\left[-\frac{1}{2}\left(\left(\theta_{i}-\mu_{i}\right)\left(\Sigma^{-1}\right)_{ij}\left(\theta_{j}-\mu_{j}\right)\right)\right], \\
    \mu_{i}                        & =\theta_{0i}+\left((H+D^{-1})^{-1}\right)_{ij}n_{j},\label{mean}                                                                                                                     \\
    \Sigma_{ij}                    & =\left((H+D^{-1})^{-1}\right)_{ij}  \quad\text{(Gaussian prior)}.\label{cov}
\end{align}
Note that the mean of the posterior distribution $\mu_{i}$ includes a term that contains the noise $\bm{n}$.
However, if we integrate over the probability distribution of the noise, this term vanishes by definition, as the noise is assumed to have zero mean.
To clarify the difference from the definition of the expected value in \cref{bayes expected value}, the expected value of the function $f(\bm{\delta t})$ of the data $\bm{\delta t}$, taken over the noise distribution, is explicitly defined as
\begin{align}
\langle f({\bm{\delta t}})\rangle_n&\equiv\int d{\bm{\delta t}}\,f({\bm{\delta t}})p({\bm{\delta t}}|\vec\theta_{0},M),
\end{align}
where the relationship between the data $\bm{\delta t}$ and the noise $\bm{n}$ is given in \cref{true data}.

Finally, consider the case where the covariance matrix $D$ of the Gaussian prior is diagonal, that is, the parameters are assumed to be uncorrelated.
It is clear that when the standard deviations of the parameters are large and $D^{-1}$ becomes negligibly small compared to the diagonal elements of $H$, the posterior distribution approaches that of the uniform prior case.
Furthermore, in the limit where the diagonal elements of the standard deviations of the parameters tend to zero, the Gaussian prior converges to a Dirac delta function, meaning that the parameter values are fixed at their true values.

\subsection{Prior}\label{subsec: Prior}

The priors are listed in \cref{Prior distributions}.
The sky locations of the pulsars ($\theta_{p}$ and $\phi_{p}$) and the standard deviation of the white noise $\sigma_{n}$ are assumed to be known.
The uniform prior is used for all parameters except for the pulsar distance.
The prior of the pulsar distance parameter is Gaussian, with a mean of 1000 pc and a standard deviation of $\sigma_{p}$ pc.
We explicitly define these priors using the Gaussian prior shown in \cref{gaussian prior}.
For parameters that originally had uniform priors, we assign large standard deviations so that the corresponding diagonal elements of the matrix $D^{-1}$ become negligible compared to those of the matrix $H$.
The covariance matrix of the Gaussian prior is explicitly given as follows:
\begin{align}
    D      & ={\rm diag}\left(\sigma'^{2}_{1},\sigma'^{2}_{2},\cdots,\sigma'^{2}_{N_{\rm others}},\sigma^{2}_{1},\sigma^{2}_{2},\cdots,\sigma^{2}_{N_{\rm psr}}\right), \\
    D^{-1} & \simeq{\rm diag}\left(0,0,\cdots,0,\frac{1}{\sigma^{2}_{1}},\frac{1}{\sigma^{2}_{2}},\cdots,\frac{1}{\sigma^{2}_{N_{\rm psr}}}\right),
\end{align}
where $\sigma'$ denotes a large standard deviation assigned to all parameters except for the pulsar distance and $N_{\rm others}$ is the number of these parameters.
The mean of the Gaussian prior consists of true values.

\begin{table}[htbp] 
    \caption{Prior distributions.}
    \label{Prior distributions}
    \centering
    \normalsize
    \begin{center}
        \begin{tabular}{lll}
            \hline
            \hline
            Parameter             & Description                                  & Prior                          \\
            \hline
            $\theta$              & Polar angle $\rm[rad]$                       & Uniform                        \\
            $\phi$                & Azimuthal angle      $\rm[rad]$              & Uniform                        \\
            $\iota$               & Inclination angle   $\rm[rad]$               & Uniform                        \\
            $\log_{10}\mathcal M$ & Logarithm of Chirp mass  $\rm[M_{\odot}]$    & Uniform                        \\
            $\log_{10}f_{0}$      & Logarithm of Initial Frequency $\rm[Hz]$     & Uniform                        \\
            $\log_{10}d_{L}$      & Logarithm of Luminosity distance $\rm[Mpc]$  & Uniform                        \\
            $\Phi_{0}$            & Initial Phase    $\rm[rad]$                  & Uniform                        \\
            $\psi$                & Polarization angle      $\rm[rad]$           & Uniform                        \\
            $L_{p}$               & Distance to the $p$-th pulsar      $\rm[pc]$ & $\mathcal{N}(1000,\sigma_{p})$ \\
            \hline
            \hline
        \end{tabular}
    \end{center}
\end{table}

Regarding the value of the standard deviation $\sigma_{p}$, four cases were considered to know the imapct of the pulsar distance determined with SKA-era precision, which are listed in \cref{distance prior}.
The parallax values used to calculate the uncertainty of the pulsar distance are summarized in \cref{distances}.
The current parallax was used in the calculations for the SKA-era.
We use the positive side of the calculated distance uncertainty for the standard deviation $\sigma_{p}$.

\begin{table}[htbp]
    \caption{Prior of pulsar distance uncertainties.}
    \label{distance prior}
    \centering
    \normalsize
    \begin{center}
        \begin{tabular}{l|llll}
            \hline
            \hline
                          & $\sigma_{p}$ for J0437-4715                         & $\sigma_{p}$ for J0030+0451 & $\sigma_{p}$ for the others \\
            \hline
            Fiducial case & \multicolumn{3}{c}{All are low precision (100 pc).}                                                             \\
            Case 1        & Current precision (1.3 pc)                          & Current precision (7.9 pc)  & Low precision (100 pc)      \\
            Case 2        & SKA-era precision (0.37 pc)                         & Current precision (7.9 pc)  & Low precision (100 pc)      \\
            Case 3        & SKA-era precision (0.37 pc)                         & SKA-era precision (1.7 pc)  & Low precision (100 pc)      \\
            \hline
            \hline
        \end{tabular}
    \end{center}
\end{table}

\begin{table}[htbp]
    \caption{Pallaxes in the present and SKA era and uncertainties in calculated pulsar distances.}
    \label{distances}
    \centering
    \normalsize
    \begin{center}
        \begin{tabular}{lllll}
            \hline
            \hline
            PSR                & Parallax [mas]     & Parallax uncertainty [mas]   & Calculated distance [pc] & Reference                                                   \\
            \hline
            Current J0437-4715  & 6.396              & $\pm 0.054$                  & $156.3{\pm 1.3}$         & \cite{dellerPRECISIONSOUTHERNHEMISPHERE2009}                \\
            Current J0030+0451 & 3.020              & $\pm 0.070$                  & $331.1^{+7.9}_{-7.5}$    & \cite{hdingMSPSRpCatalogueVLBA2023}                         \\
            SKA J0437-4715     & \multirow{2}{*}{-} & \multirow{2}{*}{$\pm 0.015$} & $156.3{\pm 0.37}$        & \multirow{2}{*}{\cite{smitsProspectsAccurateDistance2011} } \\
            SKA J0030+0451     &                    &                              & $331.1^{+1.7}_{-1.6}$    &                                                             \\
            \hline
            \hline
        \end{tabular}
    \end{center}
\end{table}

\subsection{Constant Probability Density Contour}\label{subsec: contour}

Using the closed-form posterior distribution, which is the multivariate normal distribution obtained in \cref{subsec: Closed Form Posterior}, the constant probability density contour can be calculated \cite{johnsonAppliedMultivariateStatistical2007}.
The definition of the ellipsoidal contour of the multivariate normal distribution is given in Appendix \ref{Appendix: contour}.

In particular, we compute the two-dimensional contour and its area from the bivariate normal distribution of the GW sky location ($\theta$ and $\phi$).
We consider many locations for the GW sky location of the single GW source on the celestial sphere.
We need to define the area of the two-dimensional contour of the GW sky location in order to make the area relevant for observations.
The two-dimensional contour is an ellipse and the area $S$ can be calculated as follows:
\begin{align}
    S & =\iint_{D}d\theta d\phi,\notag \\
      & =\pi a b,
\end{align}
where $D$ is the region bounded by the ellipse, $a$ is the semi-major axis and $b$ is the semi-minor axis of the ellipse.
However, $\theta$ and $\phi$ are angles in the spherical coordinate system, and it is appropriate to express the area on the unit sphere as a solid angle:
\begin{align}
    \Omega=\iint_{D}\sin\theta\,d\theta d\phi.
\end{align}
We approximate this proper area as follows:
\begin{align}
    \Omega & \simeq S\sin\theta_{0},\notag \\
           & =\pi a b\sin\theta_{0},
\end{align}
where $\theta_{0}$ is a true value of $\theta$.
This approximation would be accurate as the size of the ellipse containing $\theta_{0}$ decreases.

Finally, to compute the contour, it is necessary to calculate the mean and covariance in \cref{mean,cov}, which requires evaluating the parameter derivatives of the CGW signal in \cref{diff gw}.
We implemented the CGW signal using the Python package JAX \cite{jax2018github}, and computed its derivatives via automatic differentiation.
Recently, in the context of PTAs, the increase of observational data has led to longer computation times, and in Bayesian framework, the growing number of model parameters has further increased the time required for MCMC chains to converge.
Therefore, efficient computation of derivatives is essential, and examples of such acceleration using JAX can be found in \cite{freedmanEfficientGravitationalWave2023,vallisneriRapidParameterEstimation2024}.

\section{Result}\label{sec: Results}

We computed the contours and their areas of the uncertainty in the GW sky location ($\theta$ and $\phi$) for a single circular SMBHB using the closed-form posterior distribution.
According to Galaxy And Mass Assembly survey, the galaxy density for redshifts $z < 0.5$ (approximately 3 Gpc) is roughly estimated to be $10^{3}$ $\rm deg^{-2}$ \cite{baldryGalaxyMassAssembly2010,driverGalaxyMassAssembly2022}.
Therefore, if a localization accuracy of $10^{-3}$ $\rm deg^{2}$ $\approx 4$ $\rm amin^2$ can be achieved, it would be possible to identify the host galaxy of the GW source.
Here, as shown in \cref{parameters}, we consider cases where the GW source is located at a distance closer than 3 Gpc.
In the following, unless otherwise specified, we use the CGW parameters corresponding to the slow-evolution case described in \cref{parameters}.

\begin{figure}[htbp]
    \centering
    \includegraphics[width=16cm]{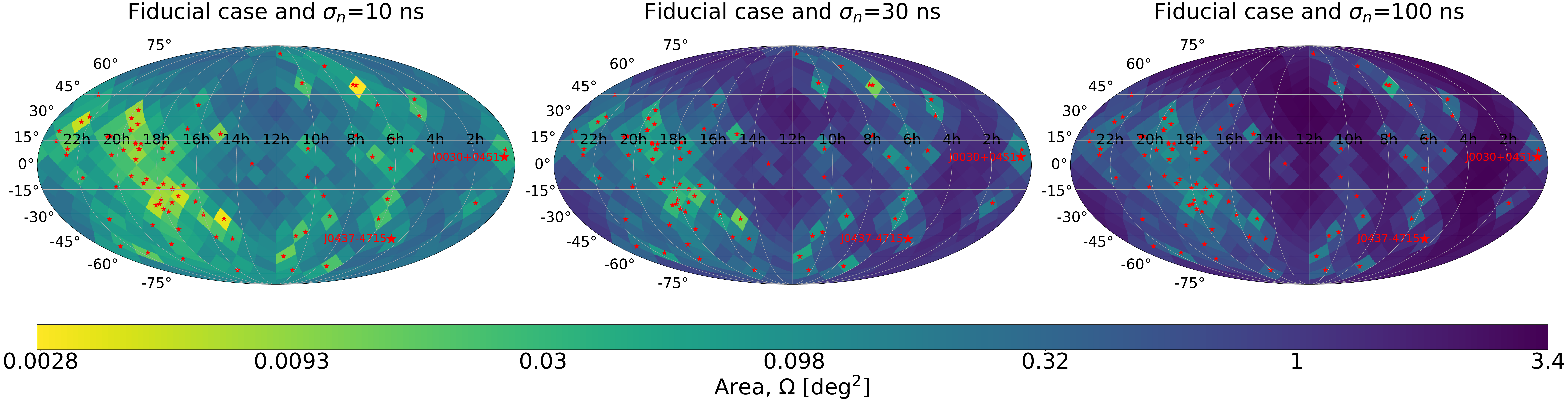}
    \caption{Areas $\Omega$ of the two-dimensional 68\% contour of the GW sky location ($\theta$ and $\phi$) as a function of the GW sky location in the ﬁducial case described in \cref{distance prior}.
        The CGW parameters used correspond to the slow evolution case listed in \cref{parameters}.
        From left to right, the standard deviations of the white noise are 10, 30, and 100 ns, respectively.
        Each GW sky location is located at the center of each pixel.
        Star markers denote the sky locations of the pulsars.
    }
    \label{fig:contour_noinfo}
\end{figure}

First, we present the results of estimating the localization accuracy of a GW source in the absence of precise pulsar distance information.
In this case, we assume an uncertainty of 100 pc for all pulsar distances, which corresponds to the fiducial case described in \cref{distance prior}.
Since this uncertainty is much larger than the GW wavelength ($\sim$ 1 pc) it provides little information for determining the phase of the pulsar term.
\cref{fig:contour_noinfo} shows the localization accuracy of GW sources as a function of the GW sky location under white noise levels of 10, 30, and 100 ns.
In all cases, the accuracy is higher (the area is smaller) in regions where pulsars are densely distributed.
It can be seen that even in the region with the best localization accuracy, $10^{-3}$ $\rm deg^{2}$ cannot be achieved.
More details regarding the dependence of localization area on white noise level are provided in Appendix \ref{Appendix: white}.

\begin{figure}[htbp]
    \centering
    \includegraphics[width=16cm]{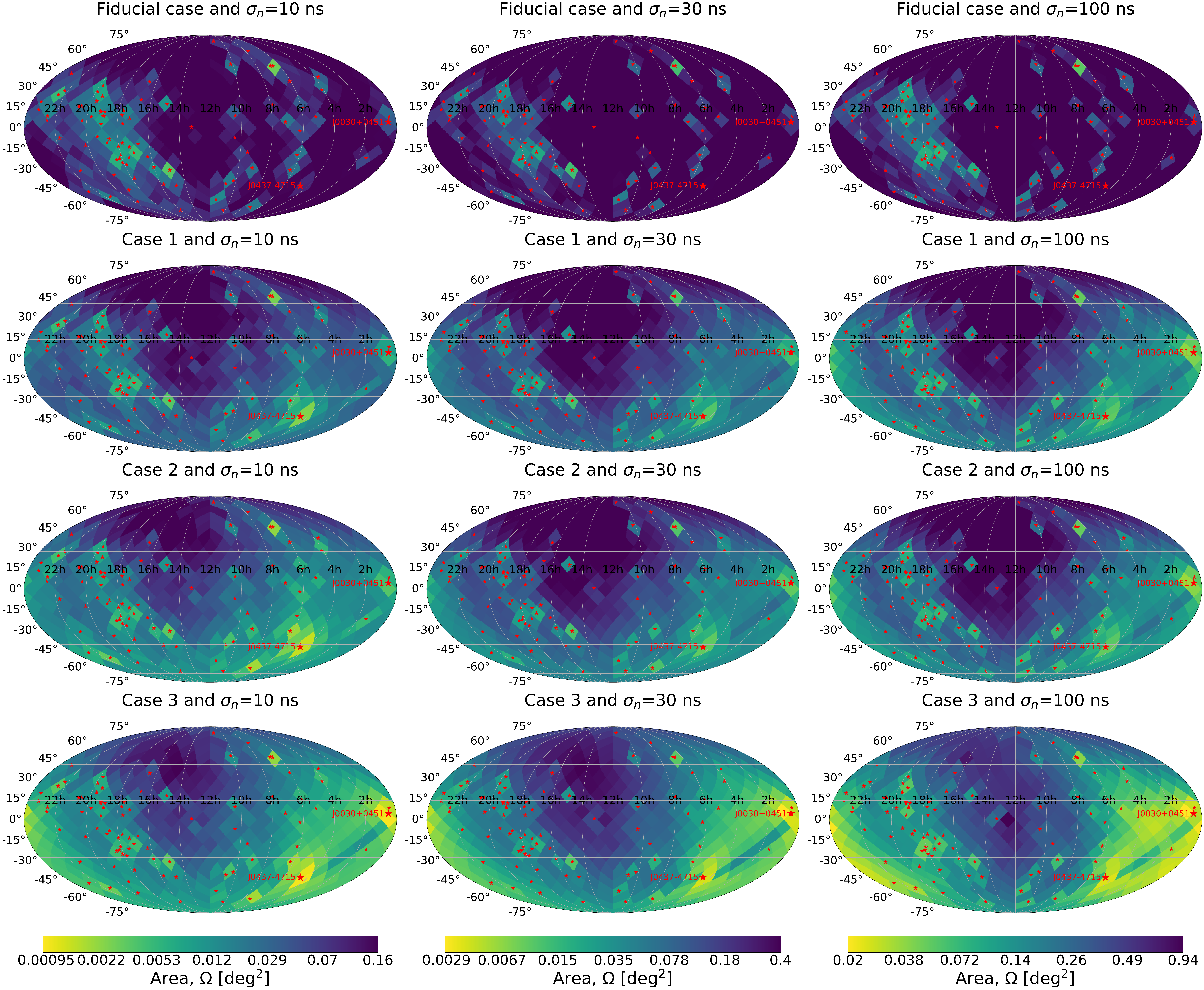}
    \caption{Same as \cref{fig:contour_noinfo}, except that from top to bottom, the rows correspond to fiducial case and Cases 1, 2 and 3 in \cref{distance prior}.
        The same color bar is used for each column.
    }
    \label{fig:contour_main}
\end{figure}

\cref{fig:contour_main} shows the localization accuracy of gravitational wave sources for white noise levels of 10, 30, and 100 ns, respectively, assuming precise distance information is available for the two pulsars J0437-4715 and J0030+0451.
Considering both the current distance precision and the expected improvements in the SKA era, we present results for the three scenarios (Cases 1, 2 and 3) described in \cref{distance prior}.
In Case 1, which uses the current distance precision for J0437-4715 and J0030+0451, the localization accuracy around these two pulsars is comparable to that in regions with dense pulsar populations in the fiducial case.
In Cases 2 and 3, which use the SKA-era precision for J0437-4715 alone and for both J0437-4715 and J0030+0451, respectively, further improvements in localization accuracy are achieved.
In particular, under Case 3 and a white noise level of 10 ns, the localization accuracy near these two pulsars reaches approximately $10^{-3}$ $\rm deg^{2}$.
Regardless of the noise level, improving the distance precision generally enhances localization accuracy.
However, the improvement around pulsars with high distance precision is nontrivial.
\cref{areas} shows the localization accuracy of GW sources within the pixels which include either J0437-4715 or J0030+0451 in \cref{fig:contour_main}.
It can be seen that whether the vicinity of J0437-4715 or J0030+0451 provides better source localization depends on the noise level.

\begin{table}[htbp]
    \caption{The localization accuracy (in $\rm deg^2$) of GW sources located within the pixels containing either J0437-4715 or J0030+0451, as shown in \cref{fig:contour_main}.}
    \label{areas}
    \centering
    \normalsize
    \begin{center}
        \begin{tabular}{l|ll|ll|ll}
            \hline
            \hline
                          & \multicolumn{2}{c|}{$\sigma_{n}=10$ ns} & \multicolumn{2}{c|}{$\sigma_{n}=30$ ns} & \multicolumn{2}{c}{$\sigma_{n}=100$ ns}                                        \\
                          & J0437-4715                             & J0030+0451                             & J0437-4715                              & J0030+0451 & J0437-4715 & J0030+0451 \\
            \hline
            Fiducial Case & 0.097                                  & 0.049                                  & 0.39                                    & 0.20        & 0.96       & 0.53       \\
            Case 1        & 0.0024                                 & 0.0058                                 & 0.011                                   & 0.013      & 0.051      & 0.042      \\
            Case 2        & 0.0014                                 & 0.0036                                 & 0.0097                                  & 0.0097     & 0.050       & 0.04       \\
            Case 3        & 0.00099                                & 0.00095                                & 0.0040                                   & 0.0039     & 0.024      & 0.032      \\
            \hline
            \hline
        \end{tabular}
    \end{center}
\end{table}

\begin{figure}[htbp]
    \centering
    \includegraphics[width=16cm]{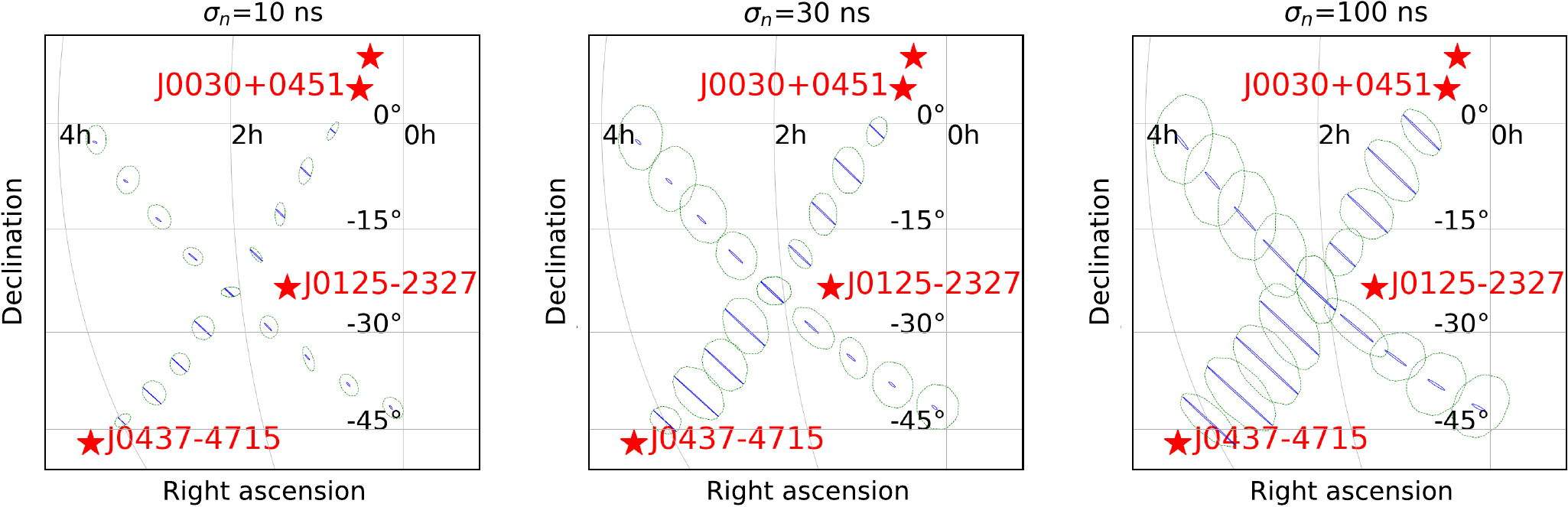}
    \caption{Contours for GW sources located between J0437-4715 and J0030+0451.
        The CGW signal parameters used are those of the slow evolution case shown in \cref{parameters}.
        The outer contour (green dashed line) corresponds to the fiducial case, while the inner contour (blue solid line) corresponds to Case 3 in \cref{distance prior}.
        The contour corresponds to the $(1 - 10^{-15}) \times 100\%$ confidence level for clarity.
    }
    \label{fig:contour_geodesic_zoom}
\end{figure}

We then focus our analysis on the regions near the two pulsars J0437-4715 and J0030+0451.
\cref{fig:contour_geodesic_zoom} illustrates the shape of the two-dimensional confidence contours for the sky location of a GW source ($\theta$ and $\phi$) when it is located between J0437-4715 and J0030+0451.
The GW sources are placed along the geodesic between the two pulsars and along its perpendicular bisector.
All adjacent GW source positions have the same angular separation.
To enhance visibility, the contours represent the $(1 - 10^{-15}) \times 100\%$ confidence level.
The outer ellipse corresponds to the fiducial case, while the inner ellipse corresponds to Case 3.
In the absence of precise distance information (fiducial case) and under 10 ns noise levels, the contours tend to be elongated along the direction of the nearest pulsar.
It should be noted that there is a third pulsar, J0125-4232, located between J0437-4715 and J0030+0451.
When SKA-era distance precision is assumed (Case 3), the uncertainty in the directions of J0437-4715 and J0030+0451 is significantly reduced.
The pulsar distance is included in $\tau_p$ in the pulsar term, as shown in \cref{timerelation}, which also depends on the inner product of the pulsar direction and the GW source direction.
This suggests that improving the pulsar-distance accuracy leads to better constraints on the angular separation between the pulsar and the GW source on the sky through $\tau_p$ in the pulsar term.

\begin{figure}[htbp]
    \centering
    \includegraphics[width=16cm]{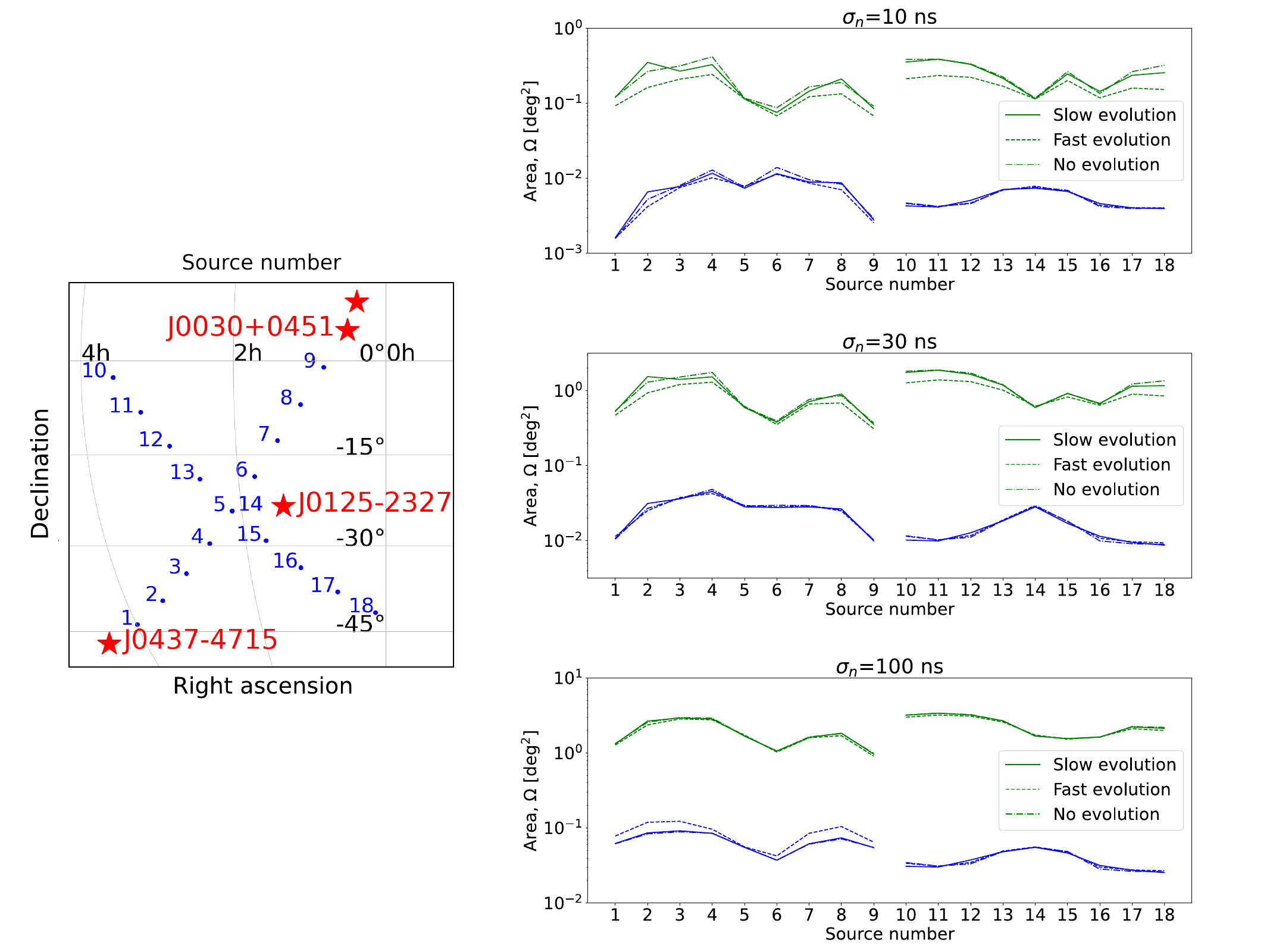}
    \caption{Left panel: The source numbers corresponding to the contours in \cref{fig:contour_geodesic_zoom}.
        Right panel: Areas $\Omega$ of the two-dimensional 68\% contours for GW sources located between J0437-4715 and J0030+0451.
        Source 14 is the same as Source 5.
        The green lines correspond to the fiducial case, while the blue lines correspond to Case 3.
        The solid, dashed, and dash-dotted lines correspond to the Slow, Fast, and No evolution cases in \cref{parameters}, respectively.
    }
    \label{fig:area_geodesic}
\end{figure}

Finally, we consider the size of the contour areas around the two pulsars.
The right panel of \cref{fig:area_geodesic} shows the areas of the contours presented in \cref{fig:contour_geodesic_zoom}, with the horizontal axis indicating the GW source numbers, as labeled in the left panel.
The comparison is made between cases without precise distance information (fiducial case) and those with SKA-era distance precision (Case 3).
Additionally, the cases with different evolutionary timescales of the supermassive black-hole binary are also examined, as listed in \cref{parameters}.
The evolutionary timescale of the binary has a relatively small impact on the localization accuracy, although a shorter timescale tends to yield slightly better precision.
First, by focusing on the GW sources labeled 10-18 located along the perpendicular bisector, we find that the pattern of area improvement differs between the fiducial case and Case 3.
In both cases, the improvement around the central GW source (labeled 14, which also corresponds to 5) is symmetric, although sources 15 and 16 are influenced by the presence of the third pulsar.
In Case 3, unlike the fiducial case, the area improvement increases with distance from the center, which will be an unexpected result.
This suggests that, for the central GW source, the axis of the contour parallel to the directions of the two pulsars has already been well constrained, and moving away from the center allows for greater improvement along the axis that has not yet been fully constrained.
Next, focusing on the GW sources located along the geodesic between the two pulsars labeled 1-9, the localization accuracy improves as the GW source is closer to the two pulsars with precise distance information.
The slight improvement in accuracy around sources 5 and 6 is attributed to the presence of the third pulsar.
Comparing Case 3 with the fiducial case, localization accuracy improves by approximately two orders of magnitude near J0437-4715, and about 1.5 orders near J0030+0451 under 10 and 30 ns noise levels.
Even between the two pulsars, and under a 100 ns noise level, more than an order of magnitude improvement is observed.

\section{Conclusions and Discussions}\label{sec: Conclusion}

We investigated the impact of precise pulsar-distance information on the localization accuracy of a single GW source in a circular orbit in the PTA experiment.
By simulating GW signals and using a closed-form expression of the posterior distribution, which is valid when the white noise is sufficiently smaller than the signal, we quantified the localization accuracy under different levels of white noise and pulsar distance precision.

According to the Galaxy and Mass Assembly survey, the number density of galaxies within 3 Gpc is typically about $10^{3}$ $\mathrm{deg^{-2}}$, implying that a localization accuracy of $10^{-3}$ $\mathrm{deg^2}$ is required to uniquely identify a host galaxy.
In practice, super-massive black-hole binaries are expected to exhibit distinctive electromagnetic signatures.
Therefore, if the candidate host galaxies can be narrowed down to several, or possibly a few dozen, it may be possible to identify the host galaxy of the gravitational-wave source through electromagnetic observations.
Then the requirement for accuracy of the angular resolution will be relaxed to $\approx 10^{-2}$ $\mathrm{deg^2}$.
However, even if the requirement is relaxed in this manner, it is unlikely that the conventional methods will achieve it, even assuming the white noise levels expected in the SKA era.
However, when precise distance information expected in the SKA-VLBI observation is given to relatively nearby pulsars such as J0437-4715 and J0030+0451, the localization accuracy improves significantly and can reach the required accuracy with the 30 ns white noise level or better.

We also demonstrated that the area and shape of the contours for the GW source localization are strongly influenced by both the geometric configuration of the PTA and the presence of pulsars with precisely measured distances.
In particular, the localization accuracy improves significantly near well-measured pulsars, with substantial enhancement also observed in the regions between two such pulsars.
Moreover, we found that slightly better localization accuracy is achieved when the GW source evolves more rapidly.

One of the limitations of this study is the assumption that the CGW signal can be described by a linearized model under the condition that the SNR is sufficiently high, referred to as the high-SNR/LSA limit.
However, this assumption does not hold for current PTA observations, which remain in the low-SNR regime.
Furthermore, recent analyses have shown that the posterior distribution can exhibit a multimodal structure in sky location \cite{falxaSearchingContinuousGravitational2023}, whereas the closed-form posterior distribution is unimodal and therefore cannot capture this feature.

As shown in Appendix \cref{fig:white_dependency}, for the slow-evolution case, the SNR of the data set is approximately 4 at a 1~$\mu$s noise level (current) and about 40 at a 100~ns noise level (future).
Therefore, although current observations do not yet achieve a sufficiently high SNR, the high-SNR/LSA limit is expected to be justified in future high-precision PTA observations.

Another limitation is that, although we performed Gaussian integration to obtain the closed-form expression, the parameters of stochastic signals cannot be treated as free parameters.
Investigating the behavior of simultaneous parameter estimation of a CGW and stochastic signals remains a subject for future work.
In particular, understanding the simultaneous parameter estimation of common-spectrum spatially uncorrelated red noise and a CGW, as demonstrated in \cite{milesMeerKATPulsarTiming2025a}, is likely to become increasingly important.
Recent studies also emphasized the importance of resolving a CGW and the stochastic GW background simultaneously \cite{ferrantiSeparatingDeterministicStochastic2024,furusawaResolvingIndividualSignal2025}.

We considered a simplified scenario in which the data contain only a continuous-wave signal and white noise.
To make the analysis more realistic, it will be necessary to account for additional components, such as the timing model, red noise, common red noise, dispersion measure variations, and Solar System ephemeris systematics \cite{falxaSearchingContinuousGravitational2023}.
It would also be interesting to apply our method to evaluate the posterior distribution for a signal from a eccentric SMBHB that require post-Newtonian correction \cite{susobhananPulsarTimingArray2020,susobhananPostNewtonianaccuratePulsarTiming2023}.

\acknowledgments
We would like to thank Polina Petrov for helpful discussions.
We are also grateful to the referee for valuable comments and suggestions.
RK is supported by JSPS KAKENHI Grant Number 24K17051.
KT is partially supported by JSPS KAKENHI Grant Numbers 20H00180, 21H01130, 21H04467, and 24H01813, Bilateral Joint Research Projects of JSPS, and the ISM Cooperative Research Program (2023-ISMCRP-2046).

\appendix

\section{Dependence of area on white noise level}\label{Appendix: white}
To investigate how the area of the source location estimate varies with the white noise level, we examine a wide range of white noise values.
For comparison with the fiducial case described in \cref{distance prior}, we also consider several different prior distributions for the pulsar distances.

\begin{figure}[htbp]
    \centering
    \includegraphics[width=8cm]{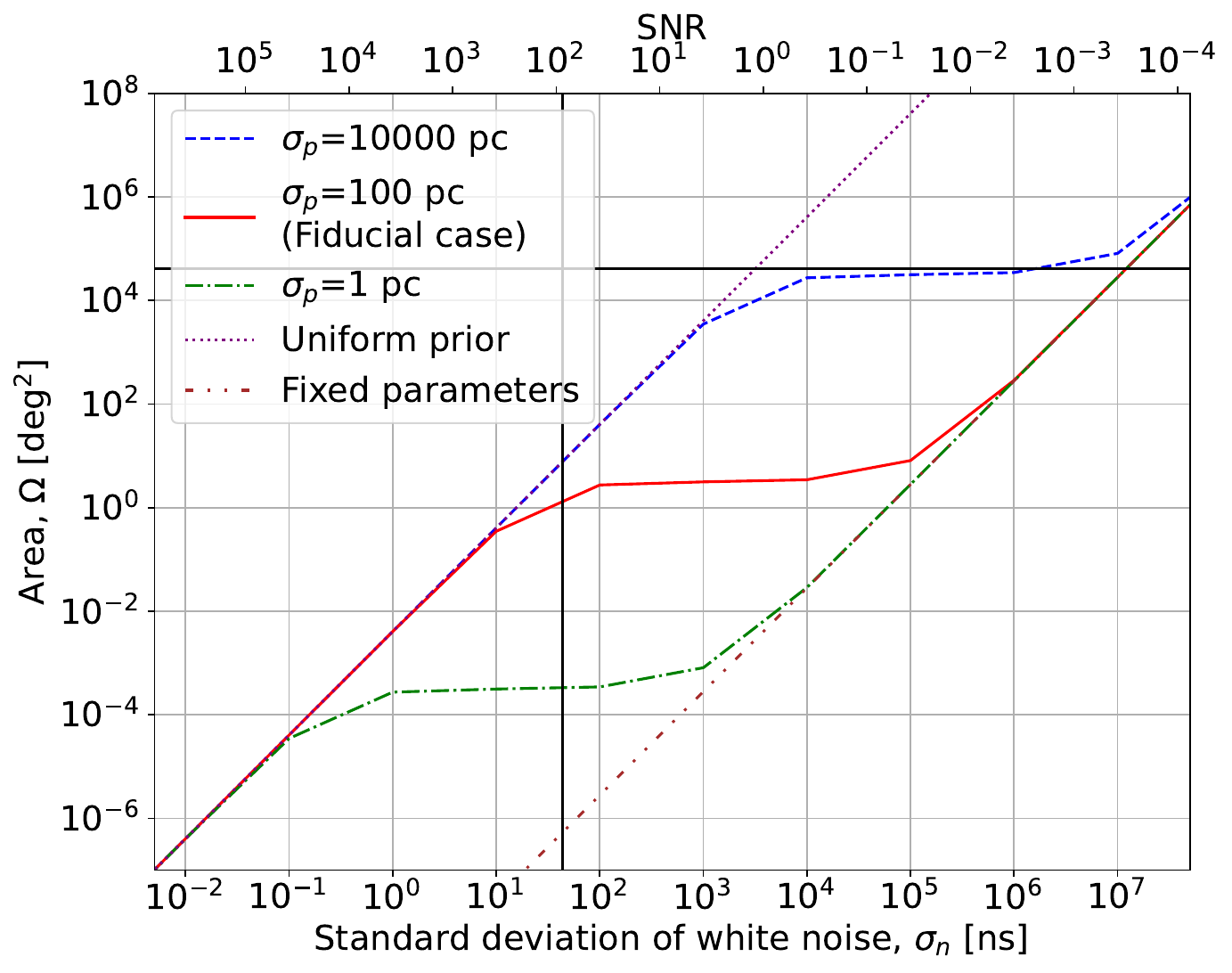}
    \caption{Plot of the area of the source location estimate versus the white noise level.
        The considered prior distributions include Gaussian priors with pulsar distance uncertainties of 100 pc (red solid line), 10000 pc (blue dashed line), and 1 pc (green dashed-dotted line), as well as a uniform prior (purple dotted line) and a fixed parameters (brown dashed-two dotted line).
        The vertical line indicates the amplitude of the Earth term.
        The horizontal line indicates the solid angle of the entire sky.
        The GW source localization is fixed to $(\theta, \phi)=(\pi/2, \pi)$.
    }
    \label{fig:white_dependency}
\end{figure}

\cref{fig:white_dependency} shows how the estimated area depends on the white noise level.
It can be seen that as the white noise level decreases, the result approaches that of the uniform prior case.
Conversely, as the noise level increases, it approaches that of the fixed parameter case.
The intermediate region between the uniform prior case and the fixed-parameter case is where the Gaussian prior exhibits nontrivial behavior.
In the fiducial case, the difference between the Gaussian prior and the uniform prior becomes apparent when the white noise exceeds 10 ns.

As a demonstration, when the uncertainty of all pulsar distances is 10000 pc, the area obtained in the intermediate region is on the same order as the solid angle of the entire sky.
Therefore, this prior does not improve the localization accuracy of the GW source.

Focusing on the uniform distribution, it can be seen that the area $\Omega$ scale as $\sigma_{n}^2$.
This leads to the relationship $\Omega \propto {\rm SNR}^{-2}$ under the assumption that the maximum likelihood estimator is approximately equal to the true value, as the optimal SNR is calculated as follows \cite{ellisPRACTICALMETHODSCONTINUOUS2012}:
\begin{align}
    {\rm SNR} & \equiv\sqrt{\bm{s}(\vec \theta_{\rm ML})^{T}\bm{C}^{-1}\bm{s}(\vec \theta_{\rm ML})},\notag \\
              & \propto \frac{1}{\sigma_{n}}, \label{snr}
\end{align}
where $\vec \theta_{\rm ML}$ is the maximum likelihood estimator.
This relationship is consistent with the result of a frequentist study \cite{sesanaMeasuringParametersMassive2010}.
It is known that, under the high-SNR/LSA limit, the covariance matrix of the posterior distribution with a uniform prior, as shown in \cref{uniform}, coincides with that of the maximum likelihood estimator \cite{rodriguezInadequaciesFisherInformation2013}.
Since the area is characterized by the covariance matrix, this consistency is an expected result.
It is also known that the mean of the maximum likelihood estimator equals the true value.
Therefore, in \cref{fig:white_dependency}, we approximated the SNR by substituting the true value in place of the estimator itself.
This SNR, evaluated using the true parameters of the signal model, corresponds to the optimal SNR of the true signal as defined in \cref{subsec: Closed Form Posterior}.
Although the difference between the maximum likelihood estimator and its mean is expected to increase with larger white noise, this effect is beyond the scope of the present analysis.

Finally, it should be noted that the high-SNR/LSA limit underlying the closed-form posterior distribution is valid only when the optimal SNR of the true signal is sufficiently high.
If the SNR becomes too low, the resulting posterior deviates from the true distribution.
For example, when the rms value of the white noise is comparable to the amplitude of the Earth term, the SNR is approximately 100 in \cref{fig:white_dependency}, which is sufficiently high to justify the use of the high-SNR/LSA limit.

\section{Contour of the multivariate normal distribution}\label{Appendix: contour}
Consider a k-dimensional normal distribution with mean $\vec\mu$ and covariance matrix $\Sigma$, whose probability density function is given by
\begin{align}
    f(\vec x)=\frac{1}{\sqrt{\det(2\pi\Sigma)}}\exp\left[-\frac{1}{2}\left(\left(\vec x-\vec \mu\right)^T \Sigma^{-1}\left(\vec x-\vec \mu\right)\right)\right].\label{normal}
\end{align}
It is known that the surface of an ellipsoid defined by the set of points $\vec x$ satisfying
\begin{align}
    \left(\vec x-\vec \mu\right)^T \Sigma^{-1}\left(\vec x-\vec \mu\right)=\chi^2_{k}(\alpha),\label{ellipse}
\end{align}
encloses $(1-\alpha)\times 100 \%$ of the total probability, where $\chi^2_{k}(\alpha)$ is the "upper" $(100 \alpha)$th percentile of the chi-square distribution with $k$ degrees of freedom \cite{johnsonAppliedMultivariateStatistical2007}.
Equivalently, $\chi^2_{k}(\alpha)$ is the value at which the cumulative distribution function (CDF) of the chi-square distribution equals $1-\alpha$.
The CDF of the chi-square distribution with two degrees of freedom takes a particularly simple form, $1-\exp(-x/2)$.
In this case, the relationship between $\chi^2_{2}(\alpha)$ and $\alpha$ is given by
\begin{align}
    1-\exp\left(-\frac{\chi^2_{2}(\alpha)}{2}\right)=1-\alpha.
\end{align}
Substituting \cref{ellipse} into \cref{normal} shows that the probability density on the surface of the ellipsoid is independent of $\vec x$ and is constant.
This surface is called a contour.
According to \cref{ellipse}, it is represented as an ellipsoid centered at $\vec\mu$, with axes given by $\pm \sqrt{\chi^2_{k}(\alpha)\lambda_{i}}\vec e_{i}$, where ($\lambda_{i}$, $\vec e_{i}$) is an eigenvalue-eigenvector pair of $\Sigma$, which satisfies $\Sigma \vec e_{i}=\lambda_{i} \vec e_{i}$ for $i=1, 2, \cdots, k$.

\bibliography{pta-concept,pta-data-release,pta-evidence,pta-cw,pta-detection,pta-residuals,pta-meerkat,pta-ska,pta-bayes,pta-fisher,pta-distance,pta-area,pta-jax,pta-ecc}

\end{document}